\documentclass[a4paper,dvips,12pt]{article}
\input{tcilatex}

\begin{document}

\title{QUANTUM CREATION OF A UNIVERSE WITH QUASI-HYPERBOLIC SPATIAL SECTIONS\thanks{%
Contribution to the XXIV National Meeting on the Physics of Particles and
Fields, held in Caxambu, MG, Brazil, in October 2003.}}
\author{Helio V. Fagundes and Teofilo Vargas \\
%EndAName
Instituto de F\'{i}sica Te\'{o}rica\\
Universidade Estadual Paulista\\
Rua Pamplona, 145, 01405-900 S\~{a}o Paulo, SP\\
Brazil}
\maketitle

\begin{abstract}
We calculate the probability of creation of a universe with space topology $%
S^{1}\times T_{g}$, where $S^{1}$ is the circle and $T_{g}$ is a compact
hyperbolic surface of genus $g\geq 2$. We use the method of path integrals
as applied to quantum cosmology.
\end{abstract}

\  \ \ 

\section{Introduction}

In recent years there has been much interest in quantum cosmology, and the
question of topology has been added to the discussion. Which is the most
probable topology of the universe? It is well known that there exist a lot
more hyperbolic manifolds than elliptic and flat ones. So it is interesting
to study the hyperbolic manifolds in quantum cosmology.

Here we shall study the probability of creation of a universe with the space
topology \newline
$S^{1}\times T_{g}$. The anisotropic minisuperspace path integral reduces to
a single ordinary integration over the lapse, then this integral is
evaluated by the lapse method. We will find that information about the space
topology is encoded in the measure of integration as well as in the
probability of creation.

The wave function of the universe as proposed by Hartle-Hawking (HH) is
given by the path integral~\cite{hh} 
\begin{equation}
\Psi (h_{ij})=\int D(g_{\mu \nu })\exp [-S_{E}(g_{\mu \nu })]  \label{wf}
\end{equation}
where $h_{ij}$ is the metric on the space section, and $S_{E}$ is the
Euclidean action of the gravitational field with a cosmological constant 
\begin{equation}
S_{E}=-\frac{1}{16\pi G}\int_{\Omega }d^{4}x\,g^{1/2}\,(R-2\Lambda )-\frac{1%
}{8\pi G}\int_{\partial \Omega }d^{3}x\,h^{1/2}\,K\,.
\end{equation}

The Euclideanized line element of the Bianchi type III universe is given by 
\begin{equation}
ds^{2} =N(\tau)^{2} d\tau^{2} + a(\tau)^{2}dr^{2} +
b(\tau)^{2}d\Omega^{2}_{g},  \label{ds}
\end{equation}
where the coordinate $r$ has a period $2\pi$, and the metric of the compact
hyperbolic surface of genus \textit{g}, $T_g$, is $d\Omega^{2}_{g}= \rho^{2}
+ \sinh^{2}{\rho}d\phi^{2}$.

The volume of the space section with the topology $S^{1}\times T_{g}$ is~
\cite{fagu} 
\begin{equation}
V=8\pi(g-1)ab^2\, ,
\end{equation}
so the Euclidean Einstein-Hilbert action is 
\begin{equation}
S_{E} =\frac{(g-1)}{2G}\int\,d\tau \left[-\frac{2\dot{a}\dot{b}b}{N}- \frac{%
a \dot{b}^{2}}{N}+Na+Nab^{2}\Lambda \right].  \label{acc}
\end{equation}
Thus the information about the topology of the spatial compact section is
encoded in the action. In the gauge in which $\dot{N}=0$, the integration of
the field equation gives 
\begin{equation}
\dot{b}^{2}+\frac{\Lambda}{3} N^{2}b^2+N^2 -\frac{C}{b}=0\,.  \label{feq}
\end{equation}
Putting the integration constant $C$ equal to zero we see that the
cosmological constant should be negative, and integrating once more, we
obtain 
\begin{equation}
b(\tau)=\sqrt{\frac{3}{|\Lambda|}}\cosh\left[\sqrt{\frac{|\Lambda|}{3}}N\tau %
\right]\,,\, a(\tau)=\sqrt{\frac{3}{|\Lambda|}}\sinh\left[\sqrt{\frac{%
|\Lambda|}{3}}N\tau \right]\,.  \label{solu2}
\end{equation}

For arbitrary $C$ the solution of equation (\ref{feq}) is expressed in terms
of elliptic funtions. The general solution for arbitrary $C$ as well as for $%
\Lambda < 0$ is under our investigation. Here we are interested in the
approximate evaluation of the path integral (\ref{wf}) by the lapse method.

\section{The lapse method and evaluation of path integral}

%%%%%%%%%%%%%%%%%%%%%%%%%%%%%%%%%%%%%%%
The lapse method, as suggested be Halliwell and Hartle~\cite{haha}, consists
in separating the anisotropic minisuperspace path integral (\ref{wf}) into
an essentially trivial functional integral over the scale factors $a(\tau)$, 
$b(\tau)$ and a nontrivial ordinary integral over the lapse $N(\tau$), and
then evaluating the propagation amplitude between fixed initial and final
values of the scale factors.

Now let us rewrite the minisuperspace action (\ref{acc}) for fixed initial
and final values of the scale factors. Rescaling the lapse $N\longrightarrow
N/a$, and expressing $S_E$ in terms of the variables $b$, $c$, and $N$, the
action is 
\begin{equation}
S_{E}(b,c,N) =\frac{(g-1)}{2G}\int\limits^{1}_{0} d\tau \left[-\frac{\dot{b}%
\dot{c}}{N}+ N+Nb^{2}\Lambda \right]\,,  \label{action}
\end{equation}
where $c=a^2b$.

The anisotropic minisuperspace quantum propagation amplitude between fixed
initial and final values of the scale factors, in the gauge $\dot{N}=0$, is
given by 
\begin{equation}
\mathbf{G}(b^{\prime \prime },c^{\prime \prime }|\,b^{\prime },c^{\prime
})=\int dN\,\int Db\,Dc\,\exp [-S_{E}(b,c,N)]\,,  \label{green}
\end{equation}
where the functional integrals over $b$ and $c$ satisfy the boundary
conditions 
\begin{equation}
b(0)=b^{\prime }\,,\,b(1)=b^{\prime \prime }\,,\,c(0)=c^{\prime
}\,,\,c(1)=c^{\prime \prime }\,.  \label{bound}
\end{equation}
The lapse funtion $N$ is in general unrestricted and then the propagator $%
\mathbf{G}$ is a solution of the Wheeler-DeWitt equation at the final state~
\cite{halou1}.

From the boundary condition we know that the classical solution must be
everywhere regular, the boundary term at $\tau=0$ should vanish, and the
quantum $4$-metric has a vanishing $3$-volume at the bottom. From (\ref
{solu2}) we see that 
\begin{equation}
a(0)=0\,, \, \frac{1}{N}\frac{da(0)}{d\tau}=1\,,  \label{coin}
\end{equation}
which is consistent with above requirement.

The initial conditions must be consistent with quantum mechanics, which
means that one is not attempting to fix too many pieces of initial data or
to fix a coordinate and its conjugate momentum simultaneously. Considering
the relationship between the velocities and the momenta in adequately chosen
new variables, and assuming $\Lambda =0$, it is shown in~\cite{halouiii}
that the wave function may be approximately evaluated by 
\begin{equation}
\Psi (a,b)\approx \int dN\,\mu (a,b,N)\exp [-S_{o}(a,b,N)]\,,  \label{hawk}
\end{equation}
where 
\begin{equation}
S_{o}(a,b,N)=\frac{(g-1)}{G}\left[ \frac{a^{4}b^{2}}{8N^{2}}-\frac{a^{2}b^{2}%
}{2N}+\frac{N}{2}\right] \,,  \label{acc3}
\end{equation}
and the measure of integration is 
\begin{equation}
\mu (a,b,N)=\frac{(g-1)}{G}f(a,b)N^{-3/2}\left[ 1-\frac{a^{2}}{2N}\right]
^{1/2}.  \label{medida}
\end{equation}
Thus the minisuperspace path integral reduces to a single ordinary
integration over the lapse.

Now let us perform a steepest descent analysis of the $N$ integration in (%
\ref{hawk}). The saddle points are the values of $N$ for which $\partial
S_{o}/\partial N=0$ and are the roots of the cubic equation, with one real $%
(N_{1})$ and a complex-conjugate pair of roots with a negative real part $%
(N_{2},N_{3})$. Analysing the asymptotic forms at large $a/b$ and ignoring
the prefactor, Halliwell and Louko~\cite{halouiii} showed that the path
integral over the real Euclidean contour with positive lapse $(N_{1})$, is
in general divergent when integrated over real Euclidean geometries, for the
half-infinite lapse contours along the positive $(N_{2})$ and negative $%
(N_{3})$ imaginary axes the semiclassical wave functions are of the form 
\begin{equation}
\Psi _{\pm }(a,b)\approx \exp \left[ \frac{(g-1)}{G}\frac{a^{2}}{8}\right]
\exp (\pm iab).
\end{equation}
These wave functions have a rapidly varying phase and a slowly varying
exponential factor, and so corresponds to an ensemble of classical
Lorentzian trajectories, weighted by the exponential factor.

Thus the (unnormalized) probability of creation of a universe with the
topology taken into consideration is 
\begin{equation}
|\Psi _{\pm }(a,b)|^{2}\approx \exp \left[ \frac{(g-1)}{G}\frac{a^{2}}{4}%
\right] \,.  \label{re}
\end{equation}

%%%%%%%%%%%%%%%%%%%%%%%%%%%%%%%%%%%%%%%%%%%%%%%%%%%%%%%%%%%%

\section{Final Remarks}

We considered the probability of creation of a universe with the space
topology $S^{1}\times T_{g}$. The anisotropic minisuperspace path integral
reduces to a single ordinary integration over the lapse, and we study this
integral by the lapse method. We found that the information about the space
topology is encoded in the measure of the integration $(\ref{medida})$ as
well as in the probability of creation (\ref{re}).

%%%%%%%%%%%%%%%%%%%%%%%%%%
\ \ \ \ 

\section*{Acknowledgment}

One of us (T. V.) would like to thank FAPESP for financial support. 
%%%%%%%%%%%%%%%%%%%%%%%%%%%

\end{document}